# A stochastic block model for community detection in attributed networks

Xiao Wang [a], Fang Dai [a,*], Wenyan Guo [a], Junfeng Wang [a]

[a] School of Science, Xi'an University of Technology, Xi'an, China

E-mail addresses: daifang@xaut.edu.cn (Dai Fang)

2210920031@stu.xaut.edu.cn (Xiao Wang)

**Abstract**—Community detection is an important content in complex network analysis. The existing community detection methods in attributed networks mostly focus on only using network structure, while the methods of integrating node attributes is mainly for the traditional community structures, and cannot detect multipartite structures and mixture structures in network. In addition, the model-based community detection methods currently proposed for attributed networks do not fully consider unique topology information of nodes, such as betweenness centrality and clustering coefficient. Therefore, a stochastic block model that integrates betweenness centrality and clustering coefficient of nodes for community detection in attributed networks, named BCSBM, is proposed in this paper. Different from other generative models for attributed networks, the generation process of links and attributes of nodes in BCSBM model follows the Poisson distribution, and the probability between community is considered based on the stochastic block model. Moreover, the betweenness centrality and clustering coefficient of nodes are introduced into the process of links and attributes of nodes generation. Finally, the expectation maximization algorithm is employed to estimate the parameters of the BCSBM model, and the node-community memberships in network is obtained through the hard division process, so the community detection is completed. By experimenting on six real-work networks containing different network structures, and comparing with the community detection results of five algorithms, the experimental results show that the BCSBM model not only inherits the advantages of the stochastic block model and can detect various network structures, but also has good data fitting ability due to introducing the betweenness centrality and clustering coefficient of nodes. Overall, the performance of this model is superior to other five compared algorithms.

**Keyworks** stochastic block model; attributed network; community detection; betweenness centrality; clustering coefficient

# 1. Introduction

An attributed network refers to a network in which the nodes or edges contain attribute information. [1].Unlike non-attributed networks that only consider the links between nodes, attributed network can model the characteristics or attributes of nodes in complex systems and provide much richer and heterogeneous information [2]. For example, in social networks, attributes can provide information about individuals' age, gender, interests, occupation, and location. In academic citation networks, attributes contain important information such as titles, authors, abstract, keywords, etc. Therefore, the studying of attributed networks is very significant to the theoretical research and application of complex systems in the real world.

Recently, the community detection in attributed networks has attracted widespread attention from researchers. Zhou et al. [4] proposed the SA-Cluster algorithm, which uses a unified distance measure to combine topology structures and attributes to construct a weighted network, and employed a clustering algorithm based on K-Medoids [5] to mine the community structure in the weighted network. This algorithms is similarity-based for community detection in attributed networks, the relevant algorithms are SAGL [6], PWMN [7], ANCA [8] and SAS-LP [9]. In addition, Wang et al. [10] proposed the SCI algorithm, which uses a two-factor non-matrix factorization method to extract the node memberships matrix from the attribute matrix and adjacency matrix to achieve community detection. The algorithm is based on non-negative matrix factorization for community detection in attributed networks, the related algorithms are PICS [11], SCD [12], MDNMF [13] and TENE [14].

These above algorithms are widely applied to detect communities with assortative structures shown as Fig. 1(a), i.e., tight intra-community node links and relatively sparse inter-community node links [15]. However, in the real world, there are not only assortative structures but also disassortative structures [16], such as multipartite structures shown as Fig. 1(b) (node links within communities are sparse, and node links between communities are reversely densely), and mixture structures such as Fig. 1(c) (It contains both community structure and multipartite structures). Currently, the main approach that can deal with both assortative networks and disassortative networks is stochastic block model (SBM) [18]. The SBM is a common model for characterizing networks with complex structural patterns, such as community, multipartite and mixture structures. Chai et al.[19] inherited the advantages of general stochastic block model (GSB) [20] and popularity and productivity link model (PPL) [21], introduced the popularity and productivity of nodes to simulate the scale-free properties of real networks and proposed PPSB_DC model. A two-stage projection algorithm is used to estimate model parameters for PPSB_DC, but a recent study found [22]: this two-stage algorithm does not guarantee convergence. Chen et al. [23] proposed BNPA model based on Newman's mixture models (NMM) [24] and Bayesian nonparametric theory. This model not only makes full use of the links between nodes and attributes of nodes to divide communities by sharing hidden variables, but also employs Bayesian nonparametric theory to determine the number of communities automatically, which

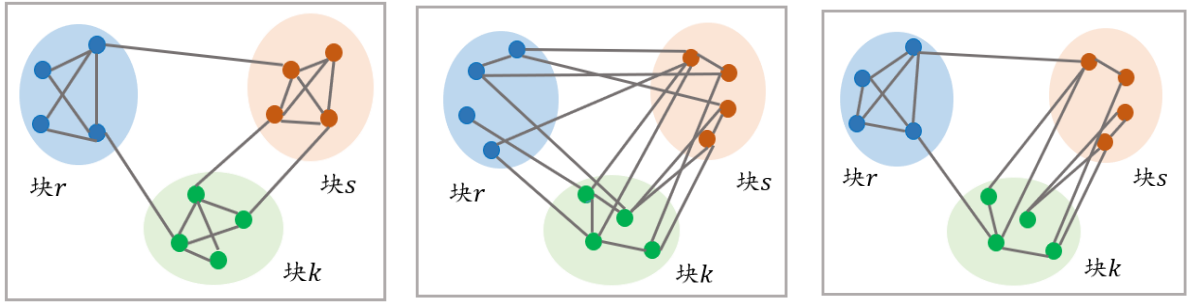

Fig.1 Examples of assortative and disassortative networks [17]

(a) assortative network with community structures; (b) disassortative network with multipartite structures; (c) disassortative network with mixture structures solves the problem of other methods that need to define the number of communities in advance, but it may be inaccurate to infer the number of communities, which will affect the accuracy of community detection. He et al. [25] proposed NEMBP model, which combines degree-corrected stochastic block model (DSBM) [26] and multinomial distributions to model the generation process of links and attributes to better fit the real network. The model parameters are inferred by using the nested Expectation-Maximum (EM) algorithm [27] and belief propagation (BP) [28]. Compared to the EM algorithm alone, the solution complexity of the NEMBP model is higher. Chen et al. [29] proposed subspace stochastic block model (SSB), which not only incorporates the attributes of nodes into the GSB model in the form of probabilities, but also constructs a hidden network by integrating the topology information and attribute information in the process of generating links between nodes. In this hidden network, it is possible to generate links between all nodes, avoiding the consideration of unobserved edges, thus avoiding the negative sampling strategy. Chang et al. [22] proposed PSB_PG model, which constructs a generative model based on the potential relationship between links and attributes of nodes. However, since real networks have scale-free property, the degree of nodes follows power-law distribution, Zheng et al. [30] introduced the degree of nodes on the basis of PSB_PG and proposed the degree-corrected stochastic block model for attributed networks (DPSB_PG). However, these two models do not fully consider the unique topological information of nodes, such as betweenness centrality and clustering coefficient, etc. According to the assumption of consistency between node attributes and topology structure, it can be seen that topological information of nodes often has a great correlation with the topology structures and can affect the links between nodes.

Based on the DPSB_PG, an attributed network stochastic block model BCSBM that integrates betweenness centrality and clustering coefficient of nodes is presented in this paper, which comprehensively considers the importance of nodes and the property of node neighborhood structure. The betweenness centrality characterizes the importance of nodes by the number of shortest paths passing through a node and also describes the influence of nodes on the flow of information on the network. The clustering coefficient describes the likelihood that neighboring nodes of individuals in the network are also neighbors of each other and is used to measure the extent of node clustering. In the BCSBM, the generation of network structures and node attributes follow the Poisson distribution and are independent of each other. It is worth noting that the BCSBM model in this paper is an extension of the DPSB_PG model, and its main difference is that the two observation variables of betweenness centrality and clustering coefficient of nodes are introduced, and the performance of the model is analyzed during the experiments.

The paper is organized as follows. In Section 2, we introduce the BCSBM model. In Section 3, the parameters estimation of the BCSBM model by the EM algorithm is described. The process of community detection based on the BCSBM model is presented in Section 4. In Section 5, we give the analysis of the community detection results. Finally, we conclude our work and discuss future in Section 6.

## 2. BCSBM Model

Let $G(V, E, X)$ denotes an undirected and unweighted attributed network, where $V = \{1, 2, \cdots, n\}$ denotes the set of $n$ nodes in the network and $E = \{e_1, e_2, \cdots, e_m\}$ denotes the set of $m$ edges in the network. If the attribute of each node is denoted by a $K$ dimension vector, the attribute matrix of all nodes can be expressed as $X = (x_{ik})_{n \times K}$, where $x_{ik} = 1$ denotes node $i$ has $k$th attribute, otherwise $x_{ik} = 0$. Typically, the adjacency matrix of an undirected unweighted network is denoted by $A = (a_{ij})_{n \times n}$, where $a_{ij} = 1$ denotes node $i$ links to node $j$, otherwise $a_{ij} = 0$. Suppose a network $G$ has $c$ different communities $V_1, V_2, \cdots, V_r$ and $V = \bigcup_{r=1}^{c} V_r$.

**2.1 A generative model for integrating node topology information**

In a standard stochastic block model [18], the stochastic block probability matrix $\Theta = (\theta_{rs})_{c \times c}$ controls the probability of generating links in network, where $\theta_{rs}$ is the connecting probability of two nodes $i \in V_r$ and $j \in V_s$ and is only related to the communities to which $i$ and $j$ belong. In the PSB_PG model [22], Chang et al. relaxed this restriction by introducing a node-community memberships matrix $D = (d_{ir})_{n \times c}$, where $d_{ir}$ is the probability that a node $i$ belongs to the $r$th community $V_r$. In the DPSB_PG model [30], Zheng et al. introduced the degree of nodes into the PSB_PG model to influence the generation of network links. In the BCSBM model of this paper, the generation of links in network is strengthened by introducing betweenness centrality and clustering coefficient of nodes to affect the distribution of community.

A link between pair of nodes in network is not only related to the node-community memberships matrix $D$, the inter-community probability matrix $\Theta$ and the degree of nodes $\Gamma = (k_i)_{n \times 1}$, but also affected by the betweenness centrality of nodes $B = (b_i)_{n \times 1}$ and clustering coefficient of nodes $M = (c_i)_{n \times 1}$. So, we introduce betweenness centrality and clustering coefficient of nodes to control the network generation process, the real network can be better fitted. Assuming that the generation of links between pairs of nodes $(x, y)$ is independent and follows the Poisson distribution, the expected number of links that nodes $i$ and node $j$ lies in

communities $V_r$ and $V_s$ is

$$\hat{l}_{ij}^{rs} = \delta_i d_{ir} \theta_{rs} \delta_j d_{js}$$

where $\delta_i = k_i + c_i + b_i$, $c_i = \frac{2l_i}{k_i(k_i-1)}$ is the clustering coefficient of node $i$ [31], $b_i = \sum_{s,t \neq i} \frac{n_{st}^i}{g_{st}}$ is the betweenness centrality of node $i$ [32], $k_i$ is the degree of node $i$, $l_i$ is the total real links number of node $i$'s neighbors, $g_{st}$ is the number of shortest paths between nodes $s$ and $t$, $n_{st}^i$ is the number of those shortest paths that include node $i$. Considering all communities, the expected total number of links between nodes $i$ and $j$ is

$$\hat{l}_{ij} = \sum_{r,s=1}^{c} \delta_i d_{ir} \theta_{rs} \delta_j d_{js}$$

where $\Theta$ is symmetrical, $\sum_{i=1}^{n} \delta_i d_{ir} = 1$ and $\sum_{r,s=1}^{c} \theta_{rs} = 1$ satisfy the normalization constraints.

Suppose the generation of links is independent and the number of links follows the Poisson distribution with mean value $\hat{l}_{ij}$, given the parameters $D, \Theta$ and observed variables $T$, the probability of generating a network is

$$P(A|D, \Theta, T) = \prod_{i,j=1, i<j}^{n} \frac{(\sum_{r,s=1}^{c} \delta_i d_{ir} \theta_{rs} \delta_j d_{js})^{a_{ij}}}{a_{ij}!} \exp\left(-\sum_{r,s=1}^{c} \delta_i d_{ir} \theta_{rs} \delta_j d_{js}\right)$$

$$\times \prod_{i=1}^{n} \frac{(\frac{1}{2} \delta_i d_{ir} \theta_{rr} \delta_i d_{ir})^{a_{ii}/2}}{(a_{ii}/2)!} \exp\left(-\frac{1}{2} \sum_{r=1}^{c} \delta_i d_{ir} \theta_{rr} \delta_i d_{ir}\right)$$

(1)

where $T = \bigcup_{i=1}^{n} \{k_i, c_i, b_i\}$ is the set of degree $k_i$, clustering coefficient $c_i$ and betweenness centrality $b_i$ of the nodes.

## 2.2 A generative model for integrating node attribute information

Generally, the attributes corresponding to each node in attributed networks are high-dimensional, and whether the nodes in the community have common attributes is sparse. If the attributes of nodes in a community are highly correlated, they will also be consistent or complementary to the network topology structures, promoting the formation of community. Therefore, the generation of node attributes follows the Poisson distribution according to Poisson' theorem [39]. Let $\phi_{rk}$ denote the probability that a community $V_r$ has the $k$th attribute, and $\Phi = (\phi_{rk})_{c \times K}$ denote community-related attributes matrix. Similarly, nodes $i$ in the community $V_r$ possessing $k$th attribute related to degree $k_i$, betweenness centrality $b_i$, clustering coefficient $c_i$, node-community memberships $d_{ir}$ and community-related attributes $\phi_{rk}$. And for this reason, the propensity of node $i$ in the community $V_r$ possessing the $k$th attribute is

$$\hat{X}_{ik}^r = \delta_i d_{ir} \phi_{rk}$$

Summing over all communities $V_r$, the mean propensity of node $i$ possessing the $k$th attribute is

$$\hat{X}_{ik} = \sum_{r=1}^{c} \delta_i d_{ir} \phi_{rk}$$

where $\sum_{i=1}^{n} d_{ir} = 1$ and $\sum_{k=1}^{K} \phi_{rk} = 1$ satisfy the normalization constraints.

According to the Poisson distribution process, given the parameter matrix $D, \Phi$ and the observed variables $T = \bigcup_{i=1}^{n}\{k_i, c_i, b_i\}$, the probability $P(X|D, \Theta, T)$ of generating node attributes in network is

$$P(X|D, \Theta, T) = \prod_{i=1}^{n} \prod_{k=1}^{K} \frac{(\sum_{r=1}^{c} \delta_i d_{ir} \phi_{rk})^{x_{ik}}}{x_{ik}!} \exp\left(-\sum_{r=1}^{c} \delta_i d_{ir} \phi_{rk}\right)$$

(2)

**2.3 Integrating node topology and attribute information**

Assuming that the generative process of the adjacency matrix $A$ and attribute matrix $X$ in network are independent of each other, the joint probability is

$$P(A, X|D, \Theta, \Phi, T) = P(A|D, \Theta, T) \times P(X|D, \Phi, T)$$

$$= \prod_{i,j=1, i<j}^{n} \frac{(\sum_{r,s=1}^{c} \delta_i d_{ir} \theta_{rs} \delta_j d_{js})^{a_{ij}}}{a_{ij}!} \exp\left(-\sum_{r,s=1}^{c} \delta_i d_{ir} \theta_{rs} \delta_j d_{js}\right)$$

$$\times \prod_{i=1}^{n} \frac{(\frac{1}{2} \sum_{r=1}^{c} \delta_i d_{ir} \theta_{rr} \delta_i d_{ir})^{\frac{a_{ii}}{2}}}{(\frac{a_{ii}}{2})!} \exp\left(-\frac{1}{2} \sum_{r=1}^{c} \delta_i d_{ir} \theta_{rr} \delta_i d_{ir}\right)$$

$$\times \prod_{i=1}^{n} \prod_{k=1}^{K} \frac{(\sum_{r=1}^{c} \delta_i d_{ir} \phi_{rk})^{x_{ik}}}{x_{ik}!} \exp\left(-\sum_{r=1}^{c} \delta_i d_{ir} \phi_{rk}\right)$$

(3)

The generative process of integrating topology information and attribute information of nodes in network can be summarized as follows.

1) Extracting nodes $i$ and $j$ from the communities $V_r$ and $V_s$ with probability $d_{ir}$ and $d_{js}$, respectively.
2) Forming a link between node $i$ and node $j$ with probability $l_{ij}$, where $l_{ij} \sim Poisson(\hat{l}_{ij})$.
3) Selecting an attribute $k$ in the community $V_r$ with probability $\phi_{rk}$.
4) Selecting an attribute $k$ for the node $i$ with probability $X_{ik}$, where $X_{ik} \sim Poisson(\hat{X}_{ik})$.

The corresponding probabilistic graph model is shown in Fig 2.

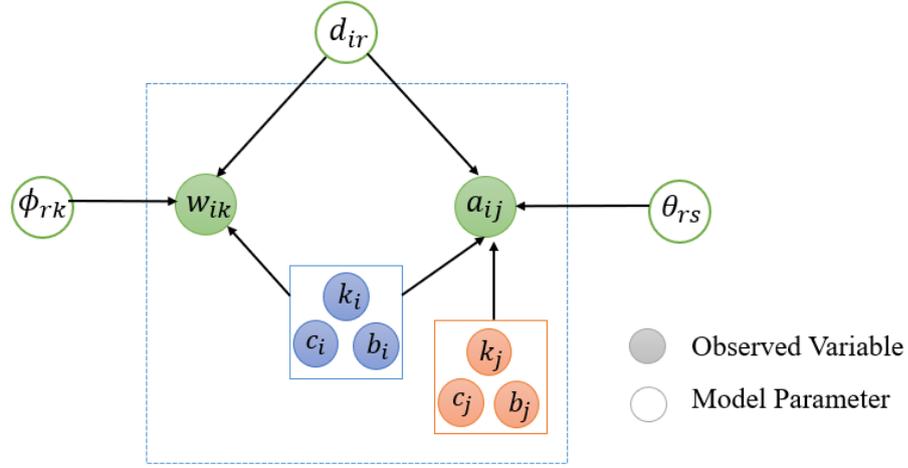

Fig.2 The probabilistic graph model for BCSBM

## 3. Estimating parameters of BCSBM model using the EM algorithm

The model BCSBM contains observed variables $A, X, T$, hidden variables $Q = (q_{ij}^{rs})_{n \times n}, \Upsilon = (\gamma_{ik}^{r})_{n \times c}$, and model parameters $D = (d_{ir})_{n \times c}$, $\Theta = (\theta_{rs})_{c \times c}$, $\Phi = (\phi_{rk})_{c \times K}$. Due to the existence of hidden variables in the model, the likelihood function cannot be solved directly, and the EM algorithm [27] is the most common hidden variable estimation method, which can handle such problems well. Therefore, in this paper, the EM algorithm is employed to estimate the parameters of the BCSBM model. The inference process is as follows.

Considering the logarithm of the Eq. (3), neglecting the constants and terms independent of model parameters, we have

$$L(D, \Theta, \Phi) = \sum_{i,j=1}^{n} \left[ \frac{1}{2} a_{ij} \ln \left( \sum_{r,s=1}^{c} \delta_i d_{ir} \theta_{rs} \delta_j d_{js} \right) - \frac{1}{2} \sum_{r,s=1}^{c} \delta_i d_{ir} \theta_{rs} \delta_j d_{js} \right]$$

$$+ \sum_{i=1}^{n} \sum_{k=1}^{K} \sum_{r=1}^{c} \left[ x_{ik} \ln \left( \sum_{r=1}^{c} \delta_i d_{ir} \phi_{rk} \right) - \sum_{r=1}^{c} \delta_i d_{ir} \phi_{rk} \right]$$

(4)

In E-step, given the parameters $D, \Theta$ and $\Phi$, and the lower bound of the log-likelihood obtained by Jensen's inequality is

$$\bar{L}(D, \Theta, \Phi) = \frac{1}{2} \sum_{i,j=1}^{n} \sum_{r,s=1}^{c} \left[ a_{ij} q_{ij}^{rs} \ln \left( \frac{\delta_i d_{ir} \theta_{rs} \delta_j d_{js}}{q_{ij}^{rs}} \right) - \delta_i d_{ir} \theta_{rs} \delta_j d_{js} \right]$$

$$+ \sum_{i=1}^{n} \sum_{k=1}^{K} \sum_{r=1}^{c} \left[ x_{ik} \gamma_{ik}^{r} \ln \left( \frac{\delta_i d_{ir} \phi_{rk}}{\gamma_{ik}^{r}} \right) - \delta_i d_{ir} \phi_{rk} \right]$$

(5)

where

$$q_{ij}^{rs} = \frac{(k_i + c_i + b_i)d_{ir}\theta_{rs}(k_j + c_j + b_j)d_{js}}{\sum_{r,s=1}^{c}(k_i + c_i + b_i)d_{ir}\theta_{rs}(k_j + c_j + b_j)d_{js}}$$

(6)

$$\gamma_{ik}^{r} = \frac{(k_i + c_i + b_i)d_{ir}\phi_{rk}}{\sum_{r=1}^{c}(k_i + c_i + b_i)d_{ir}\phi_{rk}}$$

(7)

$q_{ij}^{rs}$ denotes the probability that nodes $i$ and $j$ lie in communities $V_r$ and $V_s$, respectively, and there is a link between nodes $i$ and $j$; $\gamma_{ik}^{r}$ denotes the probability that node $i$ in the community $V_r$ and has the $k$th attribute.

In M-step, given the hidden variables $q_{ij}^{rs}$ and $\gamma_{ik}^{r}$, and we can obtain the three parameter estimates $d_{ir}, \theta_{rs}, \phi_{rk}$ according to the Lagrange multiplier method as follows

$$d_{ir} = \frac{\sum_{j=1}^{n}\sum_{s=1}^{c}a_{ij}q_{ij}^{rs} + 2 \times \sum_{k=1}^{K}x_{ik}\gamma_{ik}^{r}}{(k_i + c_i + b_i) \times \left[\sum_{i,j=1}^{n}\sum_{s=1}^{c}a_{ij}q_{ij}^{rs} + 2 \times \sum_{i=1}^{n}\sum_{k=1}^{K}x_{ik}\gamma_{ik}^{r}\right]}$$

(8)

$$\theta_{rs} = \frac{\sum_{i,j=1}^{n}a_{ij}q_{ij}^{rs}}{\sum_{i,j=1}^{n}\sum_{r,s=1}^{c}a_{ij}q_{ij}^{rs}} \qquad \phi_{rk} = \frac{\sum_{i=1}^{n}x_{ik}\gamma_{ik}^{r}}{\sum_{i=1}^{n}\sum_{k=1}^{K}x_{ik}\gamma_{ik}^{r}}$$

(9)

The derivation of the parameters $d_{ir}, \theta_{rs}, \phi_{rk}$ can be found in Appendix A. The specific steps of the parameter estimation are shown in Algorithm 1.

---

**Algorithm 1** Parameter Inference Algorithm for BCSBM

---

**Input:** the adjacency matrix $A$, the attribute matrix $X$, the number of communities $c$, the maximum iteration $I_T$ and the threshold $\epsilon$.

**Output:** the model parameters $D, \Theta, \Phi$

1: According to the adjacency matrix $A$, calculate the degree of node $k_i$, betweenness centrality of node $b_i$, and clustering coefficient of node $c_i, i = 1,2,\cdots,n$.

2: Initialize $D^{(0)}, \Theta^{(0)}, \Phi^{(0)}$.

3: Compute the objective function $L^{(0)} = (D^{(0)}, \Theta^{(0)}, \Phi^{(0)})$ by Eq. (4).

4: **for** $t = 1:I_T$ **do**

5:     E-step: Compute $q_{ij}^{rs}, \gamma_{ik}^{r}$ by Eq. (6) ~ (7). $i,j = 1,2,\cdots,n$; $r,s = 1,2,\cdots,c$.

6:     M-step: Compute $D^{(t)}, \Theta^{(t)}, \Phi^{(t)}$ by Eq. (8) ~ (9).

7:     Compute the objective function $L^{(t)} = (D^{(t)}, \Theta^{(t)}, \Phi^{(t)})$ by Eq. (4).

8:     **if** $\left|L^{(t)}(D^{(t)}, \Theta^{(t)}, \Phi^{(t)}) - L^{(t-1)}(D^{(t-1)}, \Theta^{(t-1)}, \Phi^{(t-1)})\right| < \epsilon$ or $t = I_T$ **then**

9:     $D = D^{(t)}, \Theta = \Theta^{(t)}, \Phi = \Phi^{(t)}$; STOP

10:    **end if**

11: **end for**

**Initialized scheme of $\Theta$.** In the algorithm BCSBM, the initialization of the probability matrix $\Theta$ has a great influence on the convergence speed of the algorithm. When the network structure generated by the initialization of the stochastic block probability matrix $\Theta$ is consistent with the real network structure, the algorithm will converge quickly. However, when the initial network structure (i.e., the initial values of $\Theta$) is inconsistent with the real network structure, the algorithm will converge very slowly. For the algorithm to achieve stability with as few iterations as possible, we apply maximum entropy distribution [33] and maximum likelihood to make appropriate choices for the stochastic block probability matrix $\Theta$. The specific approach is as follows [22]. The initialization of the stochastic block probability matrix $\Theta$ is divided into three schemes: (1) when the diagonal elements are larger than the non-diagonal elements, it corresponds to the assortative structures in network; (2) when the diagonal elements are smaller than the non-diagonal elements, it corresponds to disassortative structures in network; and (3) when the elements in $\Theta$ are floating around a certain value (e.g., 0.5), it corresponds to the other structures in network. The BCSBM algorithm is executed a number of times (e.g., 10 times) for each of these three cases, and then the average of the maximum likelihoods is calculated for each scheme. The initialization form of $\Theta$ corresponding to the scheme with the largest average is used as the initialization of the BCSBM algorithm.

The time complexity of BCSBM algorithm mainly depends on E-step and M-step of the EM algorithm for parameters estimation. In each iteration process, the time complexity of E-step is $O(mc^2 + nKc)$, and the time complexity of M-step is $O(nc + c^2 + nKc)$. Since the number of communities $c$ is much smaller than the number of nodes $n$, i.e., $c \ll n$. Therefore, the time complexity of M-step can be written as $O(nKc)$, and due to the maximum number of iterations of the algorithm is $I_T$, the overall time complexity of the algorithm is $O(I_T(mc^2 + nKc))$.

## 4. Community detection based on BCSBM model

Since the node-community memberships matrix $D$ characterizes the distribution of each node over all communities, our main goal is to infer the node-community memberships matrix $D = (d_{ir})_{n \times c}$, i.e., the probability that a node belongs to any community $V_r(r = 1, 2, \cdots, c)$. We infer the node-community memberships in network by hard division, that is, using $d^* = \underset{r}{\operatorname{argmax}}\{d_{ir}\}$ to limit that a node can only belong to a community. Since hidden variables are introduced when the model parameters are inferred by the EM algorithm, $D = (d_{ir})_{n \times c}$ cannot be processed directly. In order to get the hard partition, we set an operation on the parameter $D = (d_{ir})_{n \times c}, \Theta = (\theta_{rs})_{c \times c}$, that is

$$I_{ir} = \frac{\sum_{s=1}^{c} \theta_{rs} d_{ir}}{\sum_{r,s=1}^{c} \theta_{rs} d_{ir}}$$

(10)

The BCSBM algorithm can use $r^* = \underset{r}{\operatorname{argmax}}\{I_{ir}\}$ to find which community the node ultimately belongs to.

## 5. Experimental results and analysis

### 5.1 Datasets

In this paper, six real-world attributed networks are selected to examine the community detection performance of BCSBM model, including WebKB (Cornell, Texas, Washington, Wisconsin), Cora and Citeceer. The basic characteristics of the attributed networks are shown in Table 1, where $n$ and $m$ are the number of nodes and links, respectively; $K$ is the attribute type; and $c$ is the number of communities.

Table 1 Features of the attributed Networks

| Datasets | | $n$ | $m$ | $K$ | $c$ | Structure |
|---|---|---|---|---|---|---|
| WebKB | Cornell | 195 | 304 | 1703 | 5 | disassortative |
| | Texas | 187 | 328 | 1703 | 5 | disassortative |
| | Washington | 230 | 446 | 1703 | 5 | disassortative |
| | Wisconsin | 265 | 530 | 1703 | 5 | disassortative |
| Cora | — | 2708 | 5429 | 1433 | 7 | assortative |
| Citeseer | — | 3312 | 4723 | 3703 | 6 | assortative |

1) WebKB dataset [34] is a citation network consisting of web pages and links between web pages of four American universities, Cornell, Texas, Washington, and Wisconsin, with a total of 877 nodes representing all web pages, and 1,608 links representing hyperlinks between web pages. Web pages (i.e., nodes) in network are classified into the following five types, i.e., course, faculty, student, project, and staff. Each node consists of a 1703-dimensional attribute vector.
2) Cora dataset [35] is a citation network of scientific and technical literature with 2708 nodes representing all scientific publications. 5429 links representing the citation relationships from a publication to another. All the scientific publications (i.e., nodes) in network are classified into the following seven types, i.e., case-based reasoning, genetic algorithms, neural networks, probabilistic methods, reinforcement learning, rule learning, and theory. Each node consists of a 1433-dimensional attribute vector.
3) Citeseer dataset [36] is an academic citation network containing 3312 nodes representing all academic papers. 4723 links representing citation relationships between papers. All papers (i.e., nodes) in network are classified into the following six types, i.e., agents, artificial intelligence, databases, human-computer interaction, information retrieval, and machine learning. Each node consists of a 3703-dimensional attribute vector.

### 5.2 Evaluation criteria

To evaluate the community detection performance of BCSBM model on the real networks, two evaluation indexes, *Normalized Mutual Information (NMI)* and *Pairwise F-measure (PWF)*, are adopted in this paper.

(1) $NMI$

The $NMI$ proposed in [37] is based on the confusion matrix to judge the completeness of information retention after community division. Its definition is shown in Eq. (11).

$$NMI(A,B) = \frac{-2\sum_{r=1}^{c_1}\sum_{s=1}^{c_2} N_{rs} \log\left(\frac{N_{rs}N}{N_r N_s}\right)}{\sum_{r=1}^{c_1} N_r \log\left(\frac{N_r}{N}\right) + \sum_{s=1}^{c_2} N_s \log\left(\frac{N_s}{N}\right)}$$

(11)

Where $A$ is real community, $B$ is the community divided by the community detection algorithm, $c_1$ denotes the number of real community $A$, $c_2$ denotes the number of community $B$ divided by the community detection algorithm, $N$ is the total number of nodes in network $G$, $N_r$, $N_s$ denote the number of nodes in communities $r$ and $s$ respectively, $N_{rs}$ denotes the number of nodes that should belong to community $r$ but are wrongly assigned to community $s$.

The range of $NMI$ is $[0,1]$, the larger the value of $NMI$ is, the better the community detection performance. If $A = B$, $NMI(A,B) = 1$. If $A$ and $B$ are completely different, $NMI(A,B) = 0$.

(2) $PWF$

The $PWF$ proposed in [38] integrates the concepts of precision and recall into a single evaluation, and its definition is shown in Eq. (12).

$$PWF = \frac{2 \times Precision \times Recall}{Precision \times Recall}$$

(12)

where $Precision = |S \cap T|/|S|$, $Recall = |S \cap T|/|T|$ denote the precision and recall of the division results of the community detection algorithm, respectively, $S$ denotes the set of nodes that are assigned to the same community, $T$ denotes the set of nodes that have the same label, and $|\cdot|$ denotes the number of elements in the set. The range of values of $PWF$ is also $[0,1]$. The larger the value of $PWF$ is, the better the partitioning effect of community algorithm division.

### 5.3 Experimental results and analysis

To verify the validity of the BCSBM model, in this section, experiments are conducted on the six real-world attributed networks shown in Table 1 and compared with the existing generative models that integrate links and attributes of nodes, PPSB_DC [19], BNPA [23], NEMBP [25], PSB_PG [22], and DPSB_PG [30]. In order to maintain fairness, all algorithms keep the optimal parameter settings mentioned in the original paper, and the experimental results are shown in Table 2 and Table 3, Fig.3 and Fig.4. Table 2 and Fig.3 show the $NMI$ metrics of the six algorithms, and Table 3 and Fig.4 show the $PWF$ metrics of the six algorithms. Since the EM algorithm is particularly sensitive to initial values, we conducted 30 experiments on each model, and the mean and maximum values of 30 times are given for two indicators.

Table 2  *NMI* of the BCSBM model and compared algorithms on attributed networks

| Datasets | NMI Value | Model | | | | | |
|---|---|---|---|---|---|---|---|
| | | PPSB_DC | BNPA | NEMBP | PSB_PG | DPSB_PG | BCSBM |
| Cornell | mean | 0.1128 | 0.0772 | 0.1510 | 0.3131 | **0.3246** | **0.3550** |
| | max | 0.2503 | 0.0933 | 0.2793 | 0.3973 | **0.4460** | **0.4555** |
| Texas | mean | 0.2085 | 0.2265 | **0.2965** | 0.2882 | 0.2926 | **0.3214** |
| | max | 0.3663 | 0.2694 | **0.4202** | 0.3750 | 0.3933 | **0.4636** |
| Washington | mean | 0.1726 | 0.2469 | 0.1938 | **0.3235** | 0.3222 | **0.3617** |
| | max | 0.3690 | 0.2701 | 0.3107 | **0.3631** | 0.3437 | **0.4106** |
| Wisconsin | mean | 0.1315 | 0.3212 | 0.2322 | 0.3736 | **0.3772** | **0.4219** |
| | max | 0.2409 | 0.3413 | 0.4075 | 0.4230 | **0.4436** | **0.4787** |
| Cora | mean | 0.1820 | **0.4391** | **0.4033** | 0.3012 | 0.3143 | 0.3360 |
| | max | 0.5221 | **0.5022** | **0.4757** | 0.3699 | 0.3488 | 0.3593 |
| Citeseer | mean | 0.1335 | 0.1700 | 0.2003 | 0.2507 | **0.2646** | **0.3045** |
| | max | **0.3805** | 0.3196 | 0.2911 | 0.3318 | 0.3246 | **0.3862** |

Table 3  *PWF* of the BCSBM model and compared algorithms on attributed networks

| Datasets | PWF Value | Model | | | | | |
|---|---|---|---|---|---|---|---|
| | | PPSB_DC | BNPA | NEMBP | PSB_PG | DPSB_PG | BCSBM |
| Cornell | mean | 0.3789 | 0.3446 | 0.3646 | 0.4378 | **0.4498** | **0.4882** |
| | max | 0.5025 | 0.3528 | 0.4722 | 0.5672 | **0.6179** | **0.6637** |
| Texas | mean | **0.5610** | 0.5084 | **0.5622** | 0.4117 | 0.4250 | 0.4852 |
| | max | **0.6753** | 0.5257 | **0.7002** | 0.5028 | 0.5408 | 0.5854 |
| Washington | mean | 0.4751 | 0.3851 | 0.4034 | **0.4879** | 0.4829 | **0.5233** |
| | max | 0.6118 | 0.3921 | **0.6064** | 0.5483 | 0.5123 | **0.6175** |
| Wisconsin | mean | 0.3900 | 0.4818 | 0.3867 | 0.5290 | **0.5294** | **0.5916** |
| | max | 0.4839 | 0.4967 | 0.5484 | 0.5880 | **0.5953** | **0.6501** |
| Cora | mean | 0.2835 | **0.4809** | **0.4337** | 0.3554 | 0.3621 | 0.3846 |
| | max | 0.3592 | **0.5423** | **0.5203** | 0.4228 | 0.3891 | 0.4073 |
| Citeseer | mean | 0.2733 | 0.3548 | 0.3236 | 0.3561 | **0.3642** | **0.4014** |
| | max | **0.4768** | 0.3984 | 0.3804 | 0.4323 | 0.3977 | **0.4318** |

**Note 1**: The red bolded values in Table 2 and Table 3 indicate the best, and the black bolded values indicate the next best of the six models.

**Note 2**: The first 3 columns of data are from the literature [30].

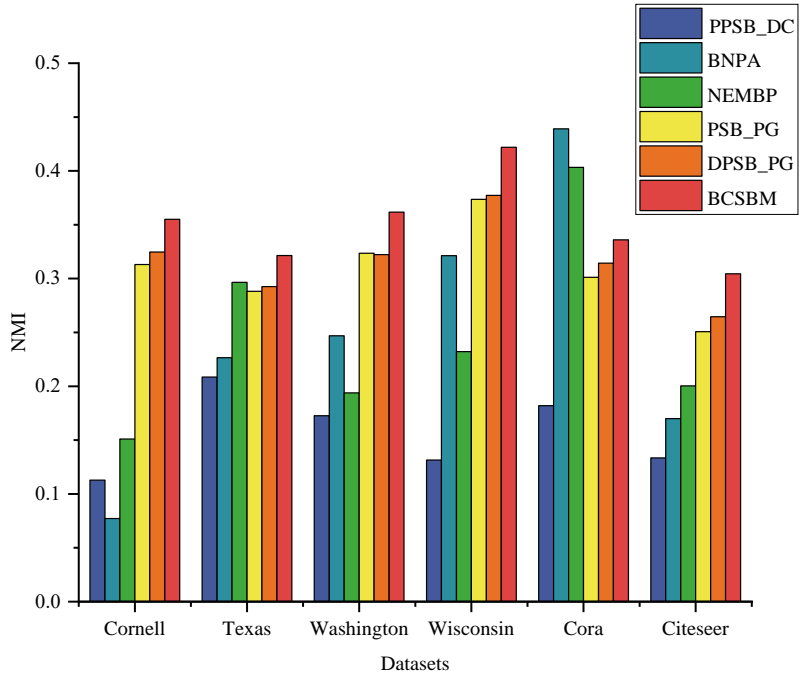

Fig.3 *NMI* of the six algorithms on attributed networks

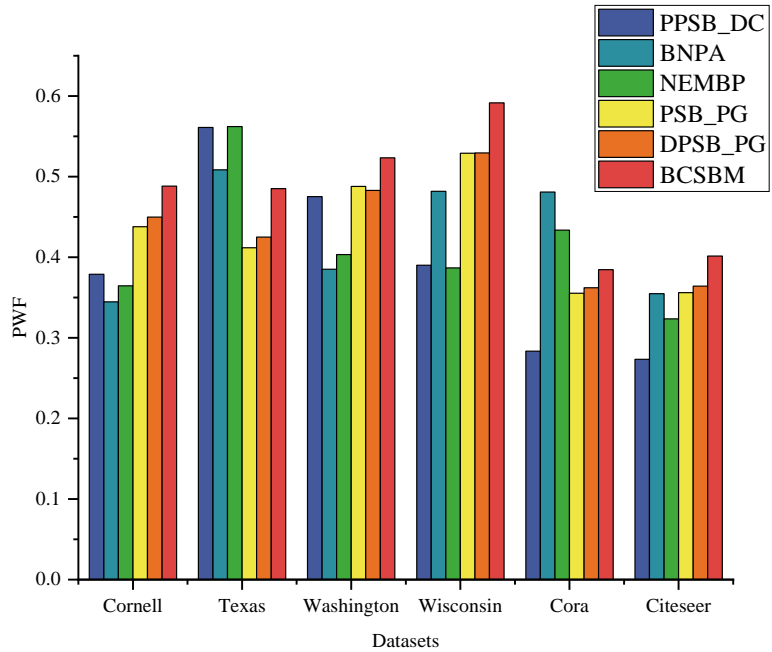

Fig.4 *PWF* of the six algorithms on attributed networks

From the above experimental results, it can be seen that the BCSBM model proposed in this paper is suitable for a variety of network structures detection, and the detection effect is significantly improved in the real attribute networks (Cornell, Texas, Washington, Wisconsin) containing disassortative structures. Compared with the PSB_PG model without considering the degree of nodes, the DPSB_PG model has improved the detection effect on attributed networks Cornell, Texas, and Wisconsin, but the performance on the Washington is slightly worse. However, the BCSBM model proposed in this paper takes into account the betweenness centrality and clustering coefficient of nodes at the same time, and performs best on the four real attribute networks with disassortative structures. The experiments show that the integration betweenness centrality and clustering coefficient of nodes have a positive impact on the community detection for attributed networks.

The BNPA model performs best on the attributed network Cora containing assortative structure, this is due to the fact that it introduces a priori information, which needs to adjust priori parameters, and its detection precision depends on the accuracy of the number of communities estimated, which does not perform well in other networks. The model BCSBM proposed in this paper performs best on the attributed network Citeseer with assortative structure, which indicates that the BCSBM model has good performance in attributed networks containing disassortative and assortative structures.

In summary, the comprehensive performance of the BCSBM model proposed in this paper is better than the other five related algorithms, and the experiments show that the stochastic block model integrating betweenness centrality and clustering coefficient of nodes, as well as the Poisson distribution can better identify the assortative and disassortative structures in attributed networks.

## 6. Conclusion and discussion

Integrating the linking relationships between nodes and the inherent attribute information of the nodes to mine the potential structure in network, and utilizing the attribute information to enhance the interpretability of the identified community, and then revealing the function of the network system is gradually being paid attention to. Based on the DPSB_PG model, the stochastic block model BCSBM that integrating betweenness centrality and clustering coefficient of nodes in attributed networks is proposed in this paper. The BCSBM model combines the network topology information and attribute information, and improves the accuracy of community detection by fitting the real network from the perspectives of node importance and node neighborhood. The uniform form of Poisson distribution facilitates the estimation of model parameters, and the EM algorithm is used to realize the parameters inference to ensure the convergence of the model. By comparing with the existing model on real attribute networks, it can be seen that BCSBM model can discover a variety of structures in networks, and the community detection accuracy is better than the DPSB_PG model, and the performance is improved in different extent compared with other existing related algorithms, which further illustrates the importance of betweenness centrality and clustering coefficient of nodes to improve the accuracy of community detection algorithm.

Since the EM algorithm may require a large amount of computation and high time complexity in the process of parameters estimation depending on the network size, it affects the efficiency of the model. Therefore, in future work, other parameter estimation methods can be considered to improve the computational efficiency of the algorithm while ensuring the accuracy of community detection.

## Acknowledgment

This work is supported by the National Natural Science Foundation of China [grant number 61976176].

## Appendix A

From Eq. (5), we know that the lower bound of the log-likelihood function is

$$\bar{L}(D, \Theta, \Phi) = \frac{1}{2} \sum_{i,j=1}^{n} \sum_{r,s=1}^{c} \left[ a_{ij} q_{ij}^{rs} \ln \left( \frac{\delta_i d_{ir} \theta_{rs} \delta_j d_{js}}{q_{ij}^{rs}} \right) - \delta_i d_{ir} \theta_{rs} \delta_j d_{js} \right]$$

$$+ \sum_{i=1}^{n} \sum_{k=1}^{K} \sum_{r=1}^{c} \left[ x_{ik} \gamma_{ik}^{r} \ln \left( \frac{\delta_i d_{ir} \phi_{rk}}{\gamma_{ik}^{r}} \right) - \delta_i d_{ir} \phi_{rk} \right]$$

(5)

Under $b_{ir} = \delta_i d_{ir}$ and $\sum_{i=1}^{n} b_{ir} = \sum_{i=1}^{n} \delta_i d_{ir} = 1$, we have

$$\tilde{L}(D) = \frac{1}{2} \sum_{i,j=1}^{n} \sum_{r,s=1}^{c} \left[ a_{ij} q_{ij}^{rs} \ln \left( \frac{b_{ir} \theta_{rs} \delta_j b_{js}}{q_{ij}^{rs}} \right) - b_{ir} \theta_{rs} b_{js} \right]$$

$$+ \sum_{i=1}^{n} \sum_{k=1}^{K} \sum_{r=1}^{c} \left[ x_{ik} \gamma_{ik}^{r} \ln \left( \frac{b_{ir} \phi_{rk}}{\gamma_{ik}^{r}} \right) - b_{ir} \phi_{rk} \right] + \sum_{r=1}^{c} \zeta_r \left( 1 - \sum_{i=1}^{n} b_{ir} \right)$$

(A1)

Taking the partial derivative of $\tilde{L}(D)$, we have

$$\frac{\partial \tilde{L}(D)}{\partial b_{ir}} = \frac{1}{2} \frac{\sum_{j=1}^{n} \sum_{s=1}^{c} a_{ij} q_{ij}^{rs}}{b_{ir}} - \frac{1}{2} \sum_{j=1}^{n} \sum_{s=1}^{c} \theta_{rs} b_{js} + \frac{\sum_{k=1}^{K} x_{ik} \gamma_{ik}^{r}}{b_{ir}} - \sum_{k=1}^{K} \phi_{rk} - \zeta_r$$

$$= \frac{1}{2} \frac{\sum_{j=1}^{n} \sum_{s=1}^{c} a_{ij} q_{ij}^{rs}}{b_{ir}} - \frac{1}{2} \sum_{r,s=1}^{c} \theta_{rs} + \frac{\sum_{k=1}^{K} x_{ik} \gamma_{ik}^{r}}{b_{ir}} - 1 - \zeta_r$$

(A2)

Let $\frac{\partial \tilde{L}(D)}{\partial b_{ir}} = 0$, then

$$\frac{1}{2} \frac{\sum_{j=1}^{n} \sum_{s=1}^{c} a_{ij} q_{ij}^{rs}}{b_{ir}} - \frac{1}{2} \sum_{r,s=1}^{c} \theta_{rs} + \frac{\sum_{k=1}^{K} x_{ik} \gamma_{ik}^{r}}{b_{ir}} - 1 - \zeta_r = 0$$

(A3)

$$\frac{1}{2} \sum_{i,j=1}^{n} \sum_{s=1}^{c} a_{ij} q_{ij}^{rs} - \frac{1}{2} \sum_{r,s=1}^{c} \theta_{rs} + \sum_{i=1}^{n} \sum_{k=1}^{K} x_{ik} \gamma_{ik}^{r} - 1 - \zeta_r = 0$$

(A4)

By Eq. (A3), Eq.(A4) and $b_{ir} = (k_i + c_i + b_i) d_{ir}$, we have $d_{ir}$ in Eq. (8) as follows.

$$d_{ir} = \frac{\sum_{j=1}^{n} \sum_{s=1}^{c} a_{ij} q_{ij}^{rs} + 2 \times \sum_{k=1}^{K} x_{ik} \gamma_{ik}^{r}}{(k_i + c_i + b_i) \times \left[\sum_{i,j=1}^{n} \sum_{s=1}^{c} a_{ij} q_{ij}^{rs} + 2 \times \sum_{i=1}^{n} \sum_{k=1}^{K} x_{ik} \gamma_{ik}^{r}\right]}$$

Note the constraint $\sum_{r,s=1}^{c} \theta_{rs} = 1$, we have

$$\tilde{L}(\Theta) = \frac{1}{2} \sum_{i,j=1}^{n} \sum_{r,s=1}^{c} \left[ a_{ij} q_{ij}^{rs} \ln\left(\frac{\delta_i d_{ir} \theta_{rs} \delta_j d_{js}}{q_{ij}^{rs}}\right) - \delta_i d_{ir} \theta_{rs} \delta_j d_{js} \right]$$
$$+ \sum_{i=1}^{n} \sum_{k=1}^{K} \sum_{r=1}^{c} \left[ x_{ik} \gamma_{ik}^{r} \ln\left(\frac{\delta_i d_{ir} \phi_{rk}}{\gamma_{ik}^{r}}\right) - \delta_i d_{ir} \phi_{rk} \right] + \mu \left(1 - \sum_{r,s=1}^{c} \theta_{rs}\right)$$

(A5)

Taking the partial derivative of $\tilde{L}(\Theta)$, we have

$$\frac{\partial \tilde{L}(\Theta)}{\partial \theta_{rs}} = \frac{1}{2} \frac{\sum_{i,j=1}^{n} a_{ij} q_{ij}^{rs}}{\theta_{rs}} - \frac{1}{2} \sum_{i,j=1}^{n} \delta_i d_{ir} \delta_j d_{js} - \mu$$

$$= \frac{1}{2} \frac{\sum_{i,j=1}^{n} a_{ij} q_{ij}^{rs}}{\theta_{rs}} - \frac{1}{2} - \mu$$

(A6)

Let $\frac{\partial \tilde{L}(\Theta)}{\partial \theta_{rs}} = 0$, then

$$\frac{1}{2} \frac{\sum_{i,j=1}^{n} a_{ij} q_{ij}^{rs}}{\theta_{rs}} - \frac{1}{2} - \mu = 0$$

(A7)

$$\sum_{i,j=1}^{n} \sum_{r,s=1}^{c} a_{ij} q_{ij}^{rs} - 1 - 2\mu = 0$$

(A8)

By Eq. (A7) and Eq.(A8), we can derive the equation $\theta_{rs}$ in Eq. (9) as follows.

$$\theta_{rs} = \frac{\sum_{i,j=1}^{n} a_{ij} q_{ij}^{rs}}{\sum_{i,j=1}^{n} \sum_{r,s=1}^{c} a_{ij} q_{ij}^{rs}}$$

Similarly, for $\phi_{rk}$, we have

$$\tilde{L}(\Phi) = \frac{1}{2} \sum_{i,j=1}^{n} \sum_{r,s=1}^{c} \left[ a_{ij} q_{ij}^{rs} \ln\left(\frac{\delta_i d_{ir} \theta_{rs} \delta_j d_{js}}{q_{ij}^{rs}}\right) - \delta_i d_{ir} \theta_{rs} \delta_j d_{js} \right]$$
$$+ \sum_{i=1}^{n} \sum_{k=1}^{K} \sum_{r=1}^{c} \left[ x_{ik} \gamma_{ik}^{r} \ln\left(\frac{\delta_i d_{ir} \phi_{rk}}{\gamma_{ik}^{r}}\right) - \delta_i d_{ir} \phi_{rk} \right] + \sum_{r=1}^{c} \xi_r \left(1 - \sum_{k=1}^{K} \phi_{rk}\right)$$

(A9)

$$\frac{\partial \tilde{L}(\Phi)}{\partial \phi_{rk}} = \frac{\sum_{i=1}^{n} x_{ik}\gamma_{ik}^{r}}{\phi_{rk}} - \sum_{i=1}^{n} \delta_i d_{ir} - \xi_r = \frac{\sum_{i=1}^{n} x_{ik}\gamma_{ik}^{r}}{\phi_{rk}} - 1 - \xi_r$$

(A10)

Let $\frac{\partial \tilde{L}(\Phi)}{\partial \phi_{rk}} = 0$, then

$$\frac{\sum_{i=1}^{n} x_{ik}\gamma_{ik}^{r}}{\phi_{rk}} - 1 - \xi_r = 0$$

(A11)

$$\sum_{i=1}^{n} \sum_{k=1}^{K} x_{ik}\gamma_{ik}^{r} - 1 - \xi_r = 0$$

(A12)

By Eq. (A11) and Eq.(A12), we have $\phi_{rk}$ in Eq. (9) in the following.

$$\phi_{rk} = \frac{\sum_{i=1}^{n} x_{ik}\gamma_{ik}^{r}}{\sum_{i=1}^{n} \sum_{k=1}^{K} x_{ik}\gamma_{ik}^{r}}$$